\documentclass[format=acmsmall, review=false, screen=true]{acmart}

\usepackage{booktabs} 

\usepackage[ruled]{algorithm2e} 
\usepackage{multirow}
\usepackage{amsmath}
\usepackage{subfig}
\usepackage{enumitem}
\usepackage{color}
\usepackage{wrapfig}

\usepackage[noend]{algpseudocode}
\usepackage{amssymb}
\usepackage{graphicx}
\usepackage{textcomp}
\usepackage[labelfont=bf]{caption}



\SetAlFnt{\small}
\SetAlCapFnt{\small}
\SetAlCapNameFnt{\small}
\SetAlCapHSkip{0pt}
\IncMargin{-\parindent}

\setcopyright{acmcopyright}
\acmJournal{TIST}
\acmYear{2019} \acmVolume{1} \acmNumber{1} \acmArticle{1} \acmMonth{1} \acmPrice{15.00}\acmDOI{1}

\received{February 2019}
\received[revised]{August 2019}
\received[accepted]{September 2019}

\begin{document}
\title{Social Science Guided Feature Engineering: A Novel Approach to Signed Link Analysis}
\author{Ghazaleh Beigi}
\affiliation{%
  \institution{Arizona State University}
  \city{Tempe}
  \state{AZ}
  \country{USA}}
\email{gbeigi@asu.edu}
\author{Jiliang Tang}
\affiliation{%
  \institution{Michigan State Universtiy}
  \city{East Lansing}
  \state{MI}
  \country{USA}
}
\email{tangjili@msu.edu}
\author{Huan Liu}
\affiliation{%
	\institution{Arizona State University}
	\city{Tempe}
	\state{AZ}
	\country{USA}}
\email{huan.liu@asu.edu}

\begin{abstract}
Many real-world relations can be represented by signed networks with positive links (e.g., friendships and trust) and negative links (e.g., foes and distrust). Link prediction helps advance tasks in social network analysis such as recommendation systems. Most existing work on link analysis focuses on unsigned social networks. The existence of negative links piques research interests in investigating whether properties and principles of signed networks differ from those of unsigned networks, and mandates dedicated efforts on link analysis for signed social networks. Recent findings suggest that properties of signed networks substantially differ from those of unsigned networks and negative links can be of significant help in signed link analysis in complementary ways. In this article, we center our discussion on a challenging problem of \emph{signed link analysis}. Signed link analysis faces the problem of data sparsity, i.e. only a small percentage of signed links are given. This problem can even get worse when negative links are much sparser than positive ones as users are inclined more towards positive disposition rather than negative. We investigate how we can take advantage of other sources of information for signed link analysis. This research is mainly guided by three social science theories, \textit{Emotional Information}, \textit{Diffusion of Innovations}, and \textit{Individual Personality}. Guided by these, we extract three categories of related features and leverage them for signed link analysis. Experiments show the significance of the features gleaned from social theories for signed link prediction and addressing the data sparsity challenge.
\end{abstract}

%
%


%
%


\maketitle

\renewcommand{\shortauthors}{G. Beigi et al.}

\section{Introduction}
The pervasive use of social media allows users to participate in online activities and produce large amounts of data. Social links are one of the most significant portion of user-generated data and can take different forms. Examples include links to users as in befriending behavior, links to entities as in purchase behavior, or links to communities as in joining behavior. Social links could be unsigned (e.g., friendship on Facebook) or signed. 
Individuals form links that represent their friendship, support and approval, or disagreement, distrust and disapproval toward each other. Positive or negative attitudes between users assign positive or negative signs to links. Examples include trust/distrust on Epinions\footnote{www.Epinions.com}, friend/foe on Slashdot\footnote{www.Slashdot.org} and vote/dispute on Wikipedia\footnote{www.Wikipedia.com}.

Positive links are important in helping users find relevant and credible information online~\cite{tang2013exploiting} and benefit many applications such as recommendation and information filtering~\cite{tang2015survey}. On the other hand, negative links could help decision makers reduce vulnerability and uncertainty associated with decision consequences~\cite{cho2006mechanism,hardin2004distrust,mcknight2001trust} and add a significant amount of knowledge than that already embedded in positive links~\cite{tang2015survey}. Link analysis is a central problem in social network analysis. It aims to understand the factors influencing the link formation~\cite{leskovec2010predicting}. Findings from link analysis have been used in a variety of social media mining tasks such as link prediction~\cite{adamic2003friends}, community detection~\cite{ruan2013efficient}, and recommendation~\cite{ma2009learning,tang2016recommendations,beigi2018similar}. For example, it has been shown that users with positive links are more likely to share similar preferences with each other than strangers and users with negative links. This indicates the importance of signed links in building effective recommendation systems~\cite{tang2016recommendations}.

Link analysis in unsigned social networks or networks with only positive links has been extensively studied. For example, users tend to form positive links with those sharing certain level of similarity with them (a.k.a. Homophily~\cite{mcpherson2001birds}), or two individuals geographically closer are more likely to connect (a.k.a. Confounding theory). Though there is a good amount of research on link analysis in unsigned social networks, it is shown that we cannot simply apply their findings to signed networks~\cite{leskovec2010predicting,tang2015survey}. For example, negative links present substantially distinct properties from positive links~\cite{szell2010multirelational}, Homophily of unsigned networks are not directly applicable to signed networks~\cite{tang2014distrust}, and signed link prediction problem differs from the unsigned link predication~\cite{beigi2016exploiting}. It is shown that negative links are not negation of positive ones; instead they have significant added value, complementary to positive ones, in the signed link analysis tasks~\cite{tang2014distrust}. For example, a small portion of negative links can improve recommendation performance~\cite{ma2009learning} or positive link prediction~\cite{guha2004propagation,leskovec2010predicting}. Moreover, signed link analysis without negative links may result in a biased estimate of positive links~\cite{tang2014distrust}. Therefore, one cannot simply extend findings of unsigned networks to signed link analysis. It is thus sensible to investigate both positive and negative links simultaneously.

Recent research on signed link analysis benefits from social psychological science theories, e.g., balance theory~\cite{heider1946attitudes,cartwright1956structural} and status theory~\cite{leskovec2010predicting}. Their success endorse signed links for various signed network mining tasks such as link prediction~\cite{leskovec2010predicting} and social psychological science theories can be used to guide link analysis in signed networks. However, majority of the existing works~\cite{naaman2018edge,javari2014cluster,shahriari2016sign,leskovec2010predicting,chiang2011exploiting} use the topological structures and properties of the existing networks, which rely on a reasonable number of signed link information. 
 As is well known, signed links in social networks are usually very sparse where most users have few in-degree or out-degree. The data sparsity for signed links gets worse as users tend to reveal more positive disposition than negative. This makes negative links much sparser than positive links. Hence, we often encounter a challenging problem of signed link sparsity.  Existing signed link analysis methods require addressing this problem. Though positive and negative links are sparse, there are substantial amounts of information that can be gleaned from users of social media networks. Some pervasively available sources of information include users' personality and emotion. Our work is to investigate if these new sources of information can be tapped on to address signed link analysis problem in face of sparsity of signed links.

Inspired by the success of link analysis with social psychological theories, we look into theories from psychology and social sciences to guide our signed link analysis research:  {\it Emotional Information}, {\it Diffusion of Innovations} and {\it Individual Personality}.  The first theory suggests that emotions of individuals toward each other are strong indicators of positive and negative links. Researchers have shown that emotions can significantly influence the level of positive and negative links between individuals~\cite{dunn2005feeling,myers2011influence,schwarz2011feelings,bodenhausen1994negative}. The second theory treats positive and negative link formation as a problem of an individual's probability of adopting a new behavior in the network following her friends' behaviors~\cite{rogers2010diffusion,valente1995network}. In our work, ``diffusion" means forming a signed link toward an individual. The third theory derived from~\cite{asendorpf1998personality,burt1998personality} suggests that personality information affects individuals' tendency to form the positive and negative links. We only consider two common types, optimism and pessimism which should be conceptualized as independent according to psychologists~\cite{herzberg2006separating,chang1998distinguishing,fischer1986optimism}. 

In our previous work~\cite{beigi2016signed}, we studied whether these social and psychological theories are applicable to user-generated data. We correlated the theories to formation of positive and negative links in signed social networks. In this work, we expand on our findings from~\cite{beigi2016signed} and seek to understand how the correlation between these theories and user-generated social media data could be utilized to compensate for data sparsity problem in signed networks. The key nuances are:
\begin{itemize}[leftmargin=*]
	\item We incorporate our findings from the above theories to guide the feature engineering for signed link analysis and extract three different categories of features from user-generated data.
	\item We deploy these social-theory-guided features for the task of link prediction in signed networks.
	\item We perform a thorough analysis and conduct extensive experiments to investigate the effectiveness of these features for the problem of signed link prediction.
	\item We study how well each category aligns with the theories. We set out to understand the importance of each category and their combination in signed link prediction. We also study how robust these social-theory-guided features are against the sparsity problem in signed link analysis.
\end{itemize}

\section{Related Work}
The ease of using the Internet has raised numerous security and privacy issues. Mitigating these concerns has been studied from different aspects such as identifying malicious activities~\cite{alvari2019hawkes,alvari2018early,alvari2017semi,alvari2019less}, addressing users' privacy issues~\cite{beigi2019identifying,beigi2018privacy,beigi2019protecting,beigi2019privacy} and studying signed links~\cite{tang2013exploiting,tang2015negative,beigi2016signed,beigi2016exploiting,beigi2014leveraging,beigi2019signed}. Link analysis~\cite{leskovec2010predicting,tang2015survey} is amongst the most popular research directions (e.g., information spread~\cite{jalili2017information}, opinion formation~\cite{jalili2013social}) to understand users' behavior in social networks. Link analysis has been extensively studied in unsigned networks~\cite{adamic2003friends}, while less effort has been devoted to signed networks~\cite{tang2013exploiting,leskovec2010predicting}. Homophily and Confounding are two examples of unsigned link analysis. Homophily suggests that users tend to create positive links with their similar peers~\cite{mcpherson2001birds}. On the other hand, Confounding considers the geographical closeness between users. Though majority of methods in signed networks have focused on only positive links (a.k.a trust prediction), recent studies have demonstrated that negative links could add value to positive ones~\cite{tang2015negative}. It is also shown that we cannot simply apply the findings of unsigned social networks to signed network link analysis~\cite{leskovec2010predicting,tang2015survey}. Thus, increasing attention has been paid to the signed link analysis~\cite{chiang2011exploiting,leskovec2010predicting}. 

Recent signed link analysis methods benefit from social psychological science theories, e.g., Homophily theory, balance theory~\cite{heider1946attitudes,cartwright1956structural} and status theory~\cite{heider1946attitudes}. Homophily is also known as assortative mixing. A network is called assortative with regard to a property if a bias is observed in favor of connections between nodes in the network with similar property~\cite{catanzaro2004social}. For example, Homophily effect suggests that similar users have a higher likelihood to establish trust relations with each other~\cite{tang2013exploiting}. Balance theory is usually bonded by phrases ``the enemy of my friend is my enemy'' and ``the friend of my enemy is my enemy''. Status theory studies the effect of nodes' ordering on signed link formation, with positive links pointing from left to right and negative links vice versa. In general, link formation works are divided into supervised and unsupervised methods. 

Supervised methods consider the positive and negative link creation problem as a classification problem by using the existing positive and negative links and train a classifier with features extracted from the signed networks\cite{chiang2011exploiting,leskovec2010predicting}. For example,~\cite{leskovec2010predicting} first extracts in-degree and out-degree numbers from positive (or negative) links and then uses balance and status theory to extract triangle-based features. Then it trains a logistic regression classifier to verify the importance of balance and status theory for positive and negative link prediction. Another work of~\cite{chiang2011exploiting}, extends the triangle-based features to the k-cycle-based features. Khodadadi et al.~\cite{khodadadi2017sign} use tendency rate of triple-micro structures in signed networks. In particular, given a pair of users, this method extracts closed triple micro structures around the given edge for two different cases: 1) the edge is positive and 2) the edge is negative. Comparing the tendency rates for two cases, the sign of the edge is determined by the scenario which has the greater tendency rate. Another work~\cite{yuan2017edge} proposes a new method to address the negative link sparsity challenge in signed link prediction problem. It first converts the original graph into the edge-dual graph. Then, instead of predicting signed links in the original graph, this method predicts sign of nodes in the edge-dual graph. This method measures the similarity between nodes using Jaccard coefficient in the new graph and then utilizes similarities to classify the sign of nodes using Support Vector Machine (SVM) model. Naaman et al.~\cite{naaman2018edge} also propose a method to assign a set of topological properties to each edge (a.k.a. VOTE) such as centrality measures, degrees, community measures and scale motif frequencies. After normalizing each attribute, a machine learning approach such as deep learning, random forest and stochastic gradient descent is applied on the extracted features to predict signs of edges.

Unsupervised methods are usually based on certain topological properties of signed networks to perform predictions~\cite{hsieh2012low,ye2013predicting}. One type is node similarity based methods~\cite{symeonidis2014transitive}, which first define similarity metrics to calculate node similarities, and then provide a way to predict the signed relations. Propagation-based methods are also used for the problem of positive and negative link formation\cite{de2006many,guha2004propagation,ziegler2005propagation}. Positive sign propagation is treated as a repeating sequence of matrix operations, which consists of four types of atomic propagations\cite{guha2004propagation}. Negative sign propagation is then incinerated after multiple steps of positive sign propagation\cite{guha2004propagation}. \cite{ziegler2005propagation} proposes to integrate negative links into the process of the Appleseed positive links computation instead of superimposing it afterwards. Another category is based on low-rank matrix factorization~\cite{hsieh2012low,ye2013predicting}. For example~\cite{hsieh2012low} mathematically models the signed link prediction problem as a low-rank matrix factorization model, based on the weak structural balance on the signed network. Also,~\cite{ye2013predicting} extends the low-rank model to perform link prediction across multiple signed networks. Another unsupervised method~\cite{javari2014cluster}, categorizes nodes into  a number of clusters so that number of negative intra-cluster links and positive inter-cluster links are minimized and clusters are balanced. It deploys a user-based collaborative filtering using similarity between clusters to predict signed edges between nodes.

Another group of works in signed network analysis is signed network embedding, which seek to capture signed network topological properties such as the variance between positive and negative links, signed link sparsity, and ratio of signed triangles. These captured properties are further used to learn a meaningful low dimensional representation for the given signed network. The final nodes' representation can be utilized for different applications such as signed link prediction~\cite{wang2017signed,derr2018signed} and node classification~\cite{perozzi2014deepwalk}. Since the focus of these works is on learning a representation for the \textit{network}, they leverage topological network structures to learn the node embeddings. Kunegis et al.~\cite{kunegis2010spectral} extend spectral analysis for signed networks. Another work proposes a matrix factorization based model which factorizes the adjacency matrix of the signed network into two low rank latent matrices and then extracts the node embeddings~\cite{hsieh2012low}. Wang et al.~\cite{wang2017signed} exploit structural balance theory and information of 2-hop networks to model the semantic meaning behind positive and negative links and extract node representations. Derr et al.~\cite{derr2018signed} also use balance theory to guide capturing important properties of the signed networks and modeling it. Another work utilizes balance theory in a graph convolutional neural networks (GCNs) model to aggregate and propagate collected signed network information across layers of the signed GCN model~\cite{derr2018signed_2}.

The vast majority of the existing algorithms utilize the topological networks structures, which relies on a reasonable number of signed link information. However, users usually establish positive links with a small proportion of users which results in sparse positive links~\cite{tang2013exploiting}. It is also significantly easier for users to express positivity than negativity in social networks and consequently negative links are often much sparser than positive links in a signed network~\cite{tang2015negative}. Therefore, the aforementioned methods suffer from the signed link sparsity problem severely. For the existing signed link analysis methods to work, it is necessary to address this sparsity problem. There are few studies~\cite{beigi2016exploiting,tang2015negative,beigi2019signed} which incorporate other available sources of information to tackle the signed link sparsity problem. For example,~\cite{tang2015negative} incorporates user interactions to predict negative links.

Exploiting user's features such as trustworthiness, bias, and optimism has been discussed in~\cite{shahriari2014ranking,shahriari2016sign,mishra2011finding,beigi2016signed,beigi2019signed}. \cite{shahriari2014ranking,shahriari2016sign} address the problem of sign prediction based on users' optimism/reputation. They define optimism as users' voting pattern and reputation as their popularity. Their approach calculates optimism as difference between number of user's positive and negative out-links. They introduce rank based optimism and reputation based on the rank of users in the signed social network. \cite{mishra2011finding} computes bias and prestige of nodes based on positive links between users in signed social networks. It defines bias as user's truthfulness. The prestige is also calculated based on opinion of other users in the form of in-links a user gets. Our work is different than~\cite{shahriari2014ranking,shahriari2016sign,mishra2011finding} as we calculate users' optimism/pessimism based on a source other than signed links, i.e. users' feedback/interactions on different entities such as items/posts. The additional sources of information could help overcome the data sparsity and imbalance problem for signed link data. 
Therefore exploiting other sources of information such as emotional and personality information can mitigate the data sparsity problem and has potentials in improving the performance of signed link prediction.
\section{Social Psychological Theories}
We aim to address the problem of sparse signed links in social networks, guided by three social psychological theories: \textit{Emotional Information}, \textit{Diffusion of Innovations} and \textit{Individual Personality}. 

\subsection{Emotional Information Theory}
Users express their emotions toward each other via various ways. In Slashdot, users comment and reply to the posts; while product-review sites such as Epinions provide the rating mechanisms for users to express their emotions toward each other. Emotional information is thus pervasively available in social media no matter how they are exposed~\cite{beigi2016exploiting}. The study of emotion exists from Aristotle era  and is based on the common sense that people have emotional experiences that are linked to their cognitive appraisal of the environment~\cite{smith1985patterns}. Emotions have shown to affect a variety of decision making processes such as the decision whether to trust/distrust a stranger (positive/negative link creation) which is likely to be influenced by person's emotional state~\cite{forgas2003affective}. According to the psychologists and sociologist, emotions of people toward each other, are strong indicators of positive and negative links. They study the impact of emotional states on positive and negative relations between individuals and support the supposition that various emotions can significantly influence the level of positive/negative relations (trust/distrust) between individuals~\cite{dunn2005feeling,myers2011influence,schwarz2011feelings,bodenhausen1994negative}. In particular, interpersonal emotions with positive valence such as happiness, gratitude and satisfaction could lead to positive links while emotions with negative valence like anger, sadness, and fear imply the negative relations. Furthermore,~\cite{bewsell2012distrust} studies the effect of negative emotions on distrust in online environments and shows that negative relations could be built when expectations are not met or negative emotions are raised. Sociologists also found that the level of emotion certainty and person's feels are the most important factors influencing the effect of an emotion on creation of positive or negative relations between people~\cite{myers2011influence}. Consequently, taking into account user's emotional information for the problem of signed link analysis, could provide a better insight.

\subsection{Diffusion of Innovation Theory}
Innovations and novel ideas do not necessarily spread at once--they instead propagate gradually through channels~\cite{young2006diffusion}. Adoption of medical and agricultural innovations are classics examples of how innovations diffuse through the society~\cite{doi:10.1086/224515,valente1995network}. Diffusion of innovation theory studies how, why and at what rate, innovations are spread among people. In particular, any idea, behavior or object that is perceived as new by the audience is considered as an innovation. Diffusion is the process of an innovation being communicated through certain channels over time among the participants of a social system~\cite{rogers2010diffusion}. Diffusion also focuses on the conditions that the likelihood of adopting a new innovation, idea or technology will decrease or increase as well as studying what qualities make innovations spread. Unlike many change theories, diffusion of innovation theory considers changes as being about reinvention of behaviors so they become a fit for individuals specific needs rather than persuading individuals to accept the change~\cite{rogers2010diffusion}.

This theory has many applications in data mining literature such as influence maximization~\cite{kempe2003maximizing} and marketing~\cite{domingos2001mining}. The problem of positive and negative link creation could be also related to the well-studied topic of {\it diffusion of innovation} by treating it as a behavior that spreads through the network. Therefore it turns into a new problem of analyzing an individual's tendency to follow her friends' behaviors toward other users. Thus, the most basic question would then be does one's probability of creating positive/negative link toward another individual depend on her friends' behavior? This question is closely related to the diffusion of innovations with a particular property that is ``diffusing" in our work is establishing a signed link toward a given individual. The answer to this question can further give insight into the problem of evolution of signed links in dynamic networks (determining who will establish a link in the future) which is not the scope of this paper.

\subsection{Individuals Personality Theory}
Users' behavior in social media could be good indicators of their personality and the reasons are two-fold~\cite{golbeck2011predicting}. First, social media allow for exposing views by providing appropriate platforms to satisfy users' basic psychological needs. Second, there is an ample amount of data regarding normative behaviors of individuals which guarantees fair analysis of their personality. Thus, rich hidden personality information available online has inspired recent studies to propose methods to extract and study them~\cite{correa2010interacts,hughes2012tale,golbeck2011predicting,seidman2013self}. As a result, exploiting user's personality information has potentials in signed link analysis. Research from sociology also suggests that people personality determines their propensity to positive/negative relations~\cite{asendorpf1998personality,burt1998personality}. Though other types of personality may exist, e.g. Big Five Model~\cite{mccrae1992introduction}, we only consider two common types, optimism and pessimism which should be conceptualized as independent according to psychologists~\cite{herzberg2006separating,marshall1992distinguishing,fischer1986optimism,chang1998distinguishing}. Note that optimism and pessimism have specific relationships with Big Five personality model dimensions~\cite{kam2012optimism,sharpe2011optimism}. In particular, optimism is broadly related to neuroticism, extroversion, agreeableness and conscientiousness. Pessimism is also strongly correlated with conscientiousness and neuroticism.

Research from  psychology and sociology suggests that people's optimism/pessimism personality implies individual's tendency to positive and negative relations~\cite{geers1998optimism,scheier1985optimism,scheier2001optimism}. 
According to Scheier et. al.~\cite{scheier1985optimism}, a person is defined as {\it optimist} when she is more likely to reinterpret negative events in a positive way and find meaning and growth in stressful situations. On the other hand, an individual is referred to as {\it pessimist} when she is pre-occupied only with the negative aspects of the environment and overlooks the positive aspects~\cite{scheier1985optimism}. Optimists have better social functioning and relations. Therefore, they actively pursue social relationships and have higher chances in establishing positive links resulting in longer lasting friendships~\cite{geers1998optimism,scheier1985optimism,Nurmi,brissette2002role,segerstrom2007optimism}. In contrast, pessimists likely practice the opposite way, i.e., having negative attitudes and expecting the worst of people and situations. Consequently, they often establish negative links with others~\cite{geers1998optimism,scheier2001optimism,Nurmi,brissette2002role,segerstrom2007optimism}. Another interesting observation of optimism/pessimism is that people generally like optimists more than pessimists and thus react more positively to optimists than to pessimists. In other words, optimists likely attract more positive links while pessimists receive more negative links~\cite{forgeard2012seeing,carver1994effects,raikkonen1999effects,helweg2002stigma,brissette2002role}. Therefore, considering user's personality information could be very helpful for studying the problem of signed link formation and signed link analysis, which leads to the issues to be discussed in the next section.

Note that these theories are different from assortative mixing concept. Assortativity concept captures bias between a pair of individuals. However, these theories are defined for an individual person and study the person's behavior toward others regardless of other people's behavior.
\section{Are Social Theories Applicable to Social Media Data?}
Before applying the above theories to signed link analysis, we first seek if these theories are applicable to social media data. We would like to first conduct a sanity check if social media data is suitable for applying these established social psychological theories. In the following, we introduce our datasets and then conduct data analysis to verify some hypotheses related to the sanity check.

\subsection{Datasets}
We collect two large online signed social networks datasets from Epinions and Slashdot where individuals can express their opinions toward each other besides creating positive and negative links\footnote{The data is available at \url{http://www.public.asu.edu/~gbeigi/TIST/}}. In addition, availability of product-rating data in Epinions and individuals' post reviews data in Slashdot, can help to approximate users' emotions toward each other. 
In contrast, personality information is not readily available online. Individuals usually do not label themselves as optimistic/pessimistic. A conventional way of obtaining personality information is to directly ask people whether they expect outcomes in their lives to be good or bad~\cite{scheier1992effects}, which is often seen in psychological surveys designed for measuring an individual's optimism and pessimism (e.g., \cite{scheier1994distinguishing}). However, since social media data is large-scale, and mainly observational, it is impractical to ask every user for their personality information. The onus is therefore on us to find a sensible way to infer if a user is optimistic/pessimistic or neither. An indirect approach is to measure optimism/pessimism based on the idea that people's expectancies for the future stem from their interpretations of the past~\cite{peterson1984causal}. Thus, past experience can reflect an individual's levels of optimism/pessimism. With social media data, the question is how to define a computational measure of optimism/pessimism. To recap, individuals do not explicitly offer their personality information, and it is infeasible to ask a large number of them about that, but individuals do leave their traces online. We ask if we can aggregate individual's data and automatically figure out if a user is optimistic or not. 

Scheier et. al. ~\cite{scheier1985optimism} defines optimism as re-interpreting negative events in a positive way and pessimism as preoccupying with the negative aspects and overlooking positive events. Following the psychology literature, user's feedback could be also used to estimate her optimism and pessimism as they are counterparts of each other~\cite{hu2013exploring,hu2014exploring}. It is shown in ~\cite{hu2013exploring,hu2014exploring} that on social media websites, optimists are more willing to give more positive feedback while pessimists are more biased toward giving more negative feedback than usual. We utilize this observation to calculate users' optimism and pessimism by leveraging their feedback to different entities in social media (e.g. items and users). Accordingly, we shall define the aforementioned aspects of personality, based on the user's item rating behavior in Epinions and opinions expressing behavior towards each other in Slashdot.

\subsubsection{Epinions} It is a product review website where users can establish trust and distrust relationships toward each other. We treat each relation as either positive or negative links and construct user-user matrix ${\bf F}$ where ${\bf F}_{ij} = 1$ if user $i$ trusts user $j$, and ${\bf F}_{ij} = -1$ if user $i$ distrusts user $j$. Also, ${\bf F}_{ij} = 0$ where the information is missing. Users can also express opinions toward each other by rating how helpful their reviews are, from 1 to 6. From these ratings, we also construct the positive and negative emotion matrices ${\bf P}$ and ${\bf N}$ as follows: (1) we consider low helpfulness ratings $\{1,2,3\}$ as negative emotions, high helpfulness ratings $\{4,5\}$ as positive emotions and the rating $3$ as neutral and (2) for each pair of users $(u_i,u_j)$, we compute the number of positive and negative emotions expressed from $u_i$ to $u_j$ to create ${\bf P}_{ij}$ and ${\bf N}_{ij}$ respectively.

We define the optimism and pessimism in Epinions as follows. Let $\mathcal{I} = \{I_1, I_2, \ldots, I_M\}$ be the set of $M$ items and assume $r_{ik}$ denotes the item rating score from $u_i$ to item $I_k$ with $r_{ik} = 0$ indicating that $u_i$ has not rated $I_k$ yet.  Also, consider $\overline{r}_k$ as the average rating score of the $k$-th item rated by users. In this paper, we consider scores in $\{1,2,3\}$ as low and $\{4,5\}$ as high scores. We use $\mathcal{O}_L(i) = \{I_k \mid r_{ik}\neq 0 \wedge \overline{r}_k\leq 3\}$ to denote the set of items with low average rating scores and rated by $u_i$.
We further use $\mathcal{O}_{HL}(i) = \{I_k \mid I_k \in  \mathcal{O}_{L}(i) \wedge {r}_{ik} > 3\}$ to denote the set of items which are scored high by $u_i$, and meanwhile have low average scores.  
Intuitively, the more frequent user $u_i$ has rated above the average, the more optimistic she is. Therefore we define the optimism score for $u_i$ as ${\bf o}_i = \frac{|\mathcal{O}_{HL}(i)|}{|\mathcal{O}_{L}(i)|}$ where $|.|$ is the size of the set.

Similarly we use $\mathcal{P}_{H}(i) = \{I_k \mid r_{ik} \neq 0 \wedge \overline{r}_k > 3\}$ to denote the set of items with high average rating scores and rated by $u_i$,
Let $\mathcal{P}_{LH}(i)$ denotes the subset of items from $\mathcal{P}_{H}(i)$, which are given low rates by $u_i$:
\begin{align}
\mathcal{P}_{LH}(i) = \{I_k \mid I_k \in  \mathcal{P}_{H}(i) \wedge {r}_{ik} \leq 3\} \nonumber
\end{align}
We define the pessimism score $u_i$ as: ${\bf p}_i = \frac{|\mathcal{P}_{LH}(i)|}{|\mathcal{P}_{H}(i)|}$. 

\subsubsection{Slashdot} It is a technology-related news platform which allows users to tag each other as either `friend' or `foe'. Similar to the Epinions, we construct user-user matrix ${\bf F}$ from the positive (friendship relations) and negative links (foes relations) in the network. Likewise, users can express their opinions and comments toward each other by annotating the articles posted by each other. In a similar way to the Epinions, using positive and negative opinions, we create user-user positive and negative emotion matrices ${\bf P}$ and ${\bf N}$ by computing the number of positive or negative emotions users express toward each other.

Additionally, we can define individual's personality in Slashdot based on user-user emotion matrices ${\bf P}$ and ${\bf N}$. Let $\overline{P}$ and $\overline{N}$ be the average of positive and negative emotions between all pairs of users, respectively. We also define $\overline{{\bf P}}_j$ and $\overline{{\bf N}}_j$ as the average of positive and negative emotions that user $u_j$ has received. Further, we define $\mathcal{O}_{L}(i)$, as a set of users $u_j$ who have received positive emotions from $u_i$, but at the same time, have received more negative emotions than the average in the network, i.e. they are worse than the average,
\begin{align}
\mathcal{O}_{L}(i) = \{u_j \mid {\bf P}_{ij}\neq 0 \wedge \overline{{\bf N}}_j> \overline{N}\} \nonumber
\end{align}
We formally define $\mathcal{O}_{HL}(i)$ to denote the set of users $u_k$ who belong to $\mathcal{O}_{L}(i)$ and have received more positive emotions from $u_i$ than $\overline{{\bf P}}_k$,
\begin{align}
\mathcal{O}_{HL}(i) = \{u_k \mid u_k \in  \mathcal{O}_{L}(i) \wedge {\bf P}_{ik} > \overline{{\bf P}}_k\} \nonumber
\end{align}
Intuitively, the more frequent $u_i$ has given positive emotions to the worst users in the network, the more optimistic she is. Therefore we define the optimism score for $u_i$ as ${\bf o}_i = \frac{|\mathcal{O}_{HL}(i)|}{|\mathcal{O}_{L}(i)|}$.

Likewise, we define $\mathcal{P}_{H}(i)$, as a set of users $u_j$ who have received negative emotions from $u_i$, but at the same time, have received more positive emotions than the average in the network, i.e. they are better than the average,
\begin{align}
\mathcal{P}_{H}(i) = \{u_j \mid {\bf N}_{ij} \neq 0 \wedge \overline{{\bf P}}_j > \overline{P}\} \nonumber
\end{align}
We define $\mathcal{P}_{LH}(i)$ to denote the set of users $u_k$ who belong to $\mathcal{P}_{H}(i)$ and have received more negative emotions from $u_i$ than $\overline{{\bf N}}_k$,
\begin{align}
\mathcal{P}_{LH}(i) = \{u_k \mid u_k \in  \mathcal{P}_{H}(i) \wedge {{\bf N}}_{ik} > \overline{N}_k\} \nonumber
\end{align}
Pessimism score of $u_i$ could be similarly defined as: ${\bf p}_i = \frac{|\mathcal{P}_{LH}(i)|}{|\mathcal{P}_{H}(i)|}$.

One thing that needs further clarification is, we shall not expect a person in real life to always behave optimistically or pessimistically-- with a high chance, they will act differently in different situations. This means, an individual could inherently possess both personality traits at very high or very low levels, but only exposes one of them at the moment~\cite{hecht2013neural}. Likewise, we allow each user to simultaneously have two personality traits with either very high or very low values.

There might be other ways to construct positive and negative emotion matrices, ${\bf P}$ and ${\bf N}$ as well as optimism and pessimism vectors, ${\bf o}$ and ${\bf p}$ such as psychological surveys which is beyond the scope of this paper. We perform some standard preprocessing in both datasets by filtering out users without both positive and negative links. Table \ref{Tab:DataStat} shows key statistics of Epinions and Slashdot.

\begin{table}[!htbp]
	\centering
	\footnotesize
	\caption{\textbf{Statistics of the preprocessed data.}}\label{Tab:DataStat}
	\vspace{-10pt}
	\begin{tabular}{l|l|l}
		& Epinions & Slashdot  \\ \hline \hline
		\# of Users & 21,308 & 6,615\\
		\# of Positive Links & 373,351 & 53,836 \\
		\# of Negative Links & 29,254 & 20,361 \\		
		\# of Positive Emotions & 8,459,770 & 870,042\\
		\# of Negative Emotions & 96,250 & 20,650\\ \hline
	\end{tabular}\vspace{-10pt}
\end{table}



\subsection{Data Analysis and Observations}
Here, we investigate how each theory is related to the formation of positive and negative links.

\subsubsection{Emotional Information Theory}
Here, we investigate (1) the existence of the correlation between emotional information and positive and negative links in signed social networks and, (2) study the impact of emotional strength on the formation of positive and negative relations. Specifically, we aim to answer the following two questions:
\begin{itemize}
	\item $\textbf{Q}_{11}$: Are users with positive (negative) emotions more likely to establish positive (negative) relations than those without? and,
	\item $\textbf{Q}_{12}$: Are users with higher positive (negative) emotion strengths more likely to create positive (negative) links than those with lower positive (negative) emotion strengths?
\end{itemize}

To answer $\textbf{Q}_{11}$, we study the relation between positive emotions and positive links. For each pair of users $(u_i, u_j)$ with positive emotions, we randomly select a user $u_k$ with no positive emotions from $u_i$ to $u_k$. We then check if positive relations from $u_i$ to $u_j$ and $u_i$ to $u_k$ exist. We set $vp = 1$ if ${\bf F}_{ij} = 1$ and $vp = 0$ otherwise. Likewise, we set $vr = 1$ if ${\bf F}_{ik}=1$ and $vr = 0$ otherwise. We obtain two vectors, ${\bf v}_p$ and ${\bf v}_r$, where ${\bf v}_p$ is the set of all $vp$s for pairs of users with positive emotions and ${\bf v}_r$ is the set of $vr$s for pairs of users without positive emotions. We conduct a one-tailed two sample $t$-test, $\{{\bf v}_p,{\bf v}_r\}$, on ${\bf v}_p$ and ${\bf v}_r$. The null and alternative hypotheses $H_0$ and $H_1$ are defined as follows:
\begin{align}
\{{\bf v}_p,{\bf v}_r\}:~~~~~~	H_0: {\bf v}_p \leq {\bf v}_r,~~~H_1: {\bf v}_p > {\bf v}_r
\end{align}

The null hypothesis $H_0$ assumes that pairs of users without positive emotions are more likely to establish positive links in comparison with pairs of users with positive emotions and the alternative hypothesis $H_1$ assumes the vice versa. The null hypothesis is rejected at significance level $\alpha = 0.01$ with p-values of $4.32\textrm{e}{-62}$ and $6.17\textrm{e}{-47}$ over Epinions and Slashdot, respectively. A similar $t$-test procedure, $\{{\bf v}_n,{\bf v}_r\}$, can be followed for negative emotions where the null hypothesis is rejected with p-values of $2.54\textrm{e}{-48}$ and $5.13\textrm{e}{-23}$ for Epinions and Slashdot, respectively. Results from $t$-tests suggest that {\it with high probability, users with positive (negative) emotions are more likely to establish positive (negative) links than those without.}

To answer $\textbf{Q}_{12}$, we rank all pairs of users $(u_i, u_j)$ with positive emotions according to their emotion strengths ${\bf P}_{ij}$ in a descending order and divide those pairs into $K$ groups $\mathcal{E} = \{\mathcal{E}_1,\mathcal{E}_2,\ldots,\mathcal{E}_K\}$ with equal sizes. The emotion strengths in $\mathcal{E}_i$ are thus larger than those in $\mathcal{E}_j$ if $i < j$. Then we form $\frac{K(K-1)}{2}$ pairs of groups $(\mathcal{E}_i,\mathcal{E}_j)$ with $i < j$ where $\mathcal{E}_i$ is the group with higher emotional strengths and $\mathcal{E}_j$ is the one with lower emotional strengths. For each pair of groups, we use $hp$ and $lp$ to denote the number of pairs of users with positive relations in $\mathcal{E}_i$ and $\mathcal{E}_j$, receptively. By repeating this over all pairs of groups, we can obtain two vectors ${\bf h}_p$ and ${\bf l}_p$ for $hp$s and $lp$s, respectively.

We conduct a one-tailed two sample $t$-test on ${\bf h}_p$ and ${\bf l}_p$ by defining the null hypothesis $H_0$: users with weak positive emotion strengths are more likely to establish positive links and the alternative one $H_1$: users with strong positive emotion strengths are more likely to create positive links:
\begin{align}
\{{\bf h}_p,{\bf l}_p\}:~~~~~~	H_0: {\bf h}_p \leq {\bf l}_p ,~~~ H_1: {\bf h}_p > {\bf l}_p.
\end{align}
By choosing $K = 10$, the null hypothesis is rejected at significance level $0.01$ with p-values of $8.47{e-23}$ and $1.72{e-19}$ for Epinions and Slashdot We make similar observations with $K = 30$ and $K = 50$. Similarly, we observe the impact of negative emotions on the formation of negative links by following a similar $t$-test on $\{{\bf h}_n,{\bf l}_n\}$. Results suggest that \textit{users with higher positive (negative) emotion strengths are more likely to establish positive (negative) links than those with lower positive emotion strengths}. Table~\ref{Tab:EmotiontTest} summarizes the p-values for the above $t$-tests at significance level $\alpha=0.01$.
\begin{table}[!htbp]
	\footnotesize
	\centering
	\caption{\textbf{P-values of $t$-test results corresponding to the emotional information theory analysis at significance level $\alpha=0.01$}. The null hypothesis for each test is refuted.}\label{Tab:EmotiontTest}
	\begin{tabular}{|l|l|l|l|l|}
		\hline
		& $\{{\bf v}_p,{\bf v}_e\}$ & $\{{\bf v}_n,{\bf v}_r\}$ & $\{{\bf h}_p,{\bf l}_p\}$ & $\{{\bf h}_n,{\bf l}_n\}$ \\ \hline
		Epinions &  $4.32\textrm{e}{-62}$ &  $2.54\textrm{e}{-48}$ &  $8.47\textrm{e}{-23}$ &  $6.12\textrm{e}{-19}$\\ \hline
		Slashdot &  $6.17\textrm{e}{-47}$ &  $5.13\textrm{e}{-23}$ &  $1.72\textrm{e}{-19}$ &  $7.28\textrm{e}{-14}$\\ \hline
	\end{tabular}\vspace{-10pt}
\end{table}

\subsubsection{Diffusion of Innovation Theory}
Following the diffusion of innovation theory, our goal here is to study if the behavior of user $u_i$ toward user $u_j$ could be influenced by the behavior of $u_i$'s friend $u_k$ toward $u_j$. More specifically, we aim to answer the following question:
\begin{itemize}
	\item $\textbf{Q}_{21}$: Is user $u_i$ with a friend $u_k$ who has a positive (negative) link to user $u_j$, more likely to establish a positive (negative) link with $u_j$ than if he/she does not have such friend?
\end{itemize}
To answer this question, we first find a pair of users $(u_i, u_j)$  where $u_i$'s friend $u_k$ has a positive link to $u_j$. We also randomly select a user $u_r$ without any positive relations with $u_k$. We then check if there are positive links from $u_i$ to $u_j$ and from $u_i$ to $u_r$. We set $fp = 1$ if ${\bf F}_{ij}=1$ and $fp = 0$ otherwise; Similarly, we set $fr = 1$ if ${\bf F}_{ir}=1$ and $fr = 0$ otherwise. We then construct two vectors, ${\bf f}_p$ and ${\bf f}_r$ where ${\bf f}_p$ is the set of all $fp$s and ${\bf f}_r$ is the set of $fr$s. We conduct a one-tailed two sample $t$-test $\{{\bf f}_p,{\bf f}_r\}$ on ${\bf f}_p$ and ${\bf f}_r$ with the null and alternative hypotheses $H_0$ and $H_1$ defined as follows:
\begin{align}
\{{\bf f}_p,{\bf f}_r\}:~~~~~~	H_0: {\bf f}_p \leq {\bf f}_r,~~~H_1: {\bf f}_p > {\bf f}_r
\end{align}

\begin{figure}[ht]
	\centering
	\footnotesize
	{\includegraphics[width=0.3\textwidth]{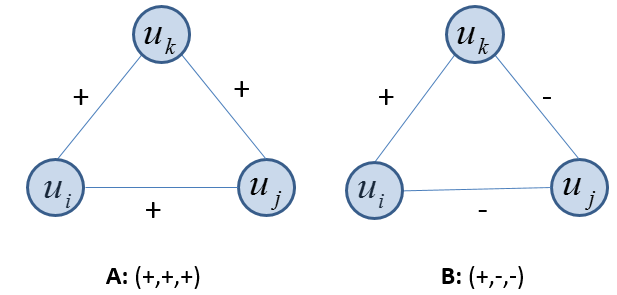}}
	\caption{\textbf{Balanced triads encouraged by diffusion of innovation theory.}}\label{fig:BalanceTriads}\vspace{-10pt}
\end{figure}

The null hypothesis is rejected at significance level $\alpha = 0.01$ with p-values of $1.84\textrm{e}{-75}$ and $3.56\textrm{e}{-91}$ over Epinions and Slashdot, respectively. Likewise, we repeat the $t$-test process $\{{\bf f}_n,{\bf f}_r\}$ for friends with negative links; however for brevity we omit the details and directly give the suggestions from the results of the one-tailed two sample $t$-test as follows: {\it users are likely to follow their friends' behaviors in terms of positive and negative link creation.} P-values for the above $t$-tests at significance level $\alpha=0.01$ are summarized in Table~\ref{Tab:DiffusiontTest}. The theory is likely to encourage triads as shown in Fig.~\ref{fig:BalanceTriads}, which are balanced according to balance theory.
\begin{table}[!htbp]
	\footnotesize
	\centering
	\vspace{-10pt}
	\caption{\textbf{P-values of $t$-test results corresponding to the diffusion of innovation theory analysis at significance level $\alpha=0.01$}. The null hypothesis for each test is refuted.}\label{Tab:DiffusiontTest}
	\begin{tabular}{|l|l|l|}
		\hline
		& $\{{\bf f}_p,{\bf f}_r\}$ & $\{{\bf f}_n,{\bf f}_r\}$ \\ \hline
		Epinions &  $1.84\textrm{e}{-75}$ &  $4.83\textrm{e}{-69}$\\ \hline
		Slashdot &  $3.56\textrm{e}{-91}$ &  $4.27\textrm{e}{-86}$\\ \hline
	\end{tabular}
\end{table}

\subsubsection{Individuals Personality Theory}\label{section:IP}
According to~\cite{golbeck2011predicting}, people's behavior observed in social media could be indicative of their personality. This is because first, social media websites allow for free interaction and exposing viewpoints by providing an appropriate platform to satisfy users' basic needs. Second, there is an ample amount of data regarding normative behaviors of individuals for analysis of individual's personality. Thus, rich hidden personality information available on social media has been used in recent studies~\cite{correa2010interacts,hughes2012tale,golbeck2011predicting,seidman2013self}.

Moreover, research from psychology and sociology suggest that people personality determines individuals' propensity to positive and negative relations~\cite{asendorpf1998personality,burt1998personality}. For example, optimistic users (1) give better responses to difficulties, (2) are happier with their lives, (3) are grateful, and (4) receive more social support. Therefore, they have higher chances in establishing and receiving positive links. In contrast, pessimists likely (1) have negative attitudes, (2) expect the worst of people and situations, and (3) treat positive events as flukes and believe that they will not happen again. Consequently, these individuals often receive or give negative links. Hence, considering users' personality information could be helpful for studying the problem of positive and negative link formation. 
Although there are many types of personalities~\cite{mccrae1992introduction}, in this paper, as a starter, we only consider two common ones, i.e., optimism and pessimism.  

Next, we investigate the impact of user's personality on the formation of positive and negative links via studying the correlation between personality information and positive and negative links. We seek to answer the following questions:
\begin{itemize}[leftmargin=*]
	\item $\textbf{Q}_{31}$: Are users with higher optimism more likely to establish positive links than those with lower optimism? and
	\item $\textbf{Q}_{32}$: Are users with higher pessimism more likely to create negative links than those with lower pessimism?
\end{itemize}

To answer the question $\textbf{Q}_{31}$, we rank all users in a descending order according to their optimism scores and divide them into $K$ levels with equal sizes denoted as $G=\{g_1, g_2,..., g_K\}$. There are $\frac{K(K-1)}{2}$ pairs of $(g_i,g_j)$ where $i<j$. We consider $g_i$ as the group of more optimistic users compared to those in $g_j$. For each pair $(g_i,g_j)$, we use $H_o$ and $L_o$ to denote the number of positive links established by users in groups $g_i$ and $g_j$, respectively. Therefore, we have two vectors ${\bf h}_o$ and ${\bf l}_o$ for $H_o$s and $L_o$s of all pairs of groups.

We conduct a one-tailed two sample $t$-test on ${\bf h}_o$ and ${\bf l}_o$ where the null hypothesis $H_0$ is that users who are less optimistic are more likely to establish positive links and the alternative hypothesis $H_1$ is that users with higher level of optimism are more likely to create positive relations:
\begin{align}
\{{\bf h}_o,{\bf l}_o\}:~~~~~~	H_0: {\bf h}_o \leq {\bf l}_o ,~~~ H_1: {\bf h}_o > {\bf l}_o.
\end{align}
We set $K = 20$, and the null hypothesis is rejected at significance level $0.01$ with p-values $3.16{e-19}$ and $1.60{e-23}$ for Epinions and Slashdot datasets, respectively. We make similar observations by setting $K = 30$ and $K = 50$. These results suggest that \textit{users with high optimistic behavior are more likely to establish positive links than those with low optimism}. To answer the question $\textbf{Q}_{32}$, we follow a similar procedure and apply $t$-test $\{{\bf h}_p,{\bf l}_p\}$ to observe the impact of pessimism on the formation of negative links. The results suggest that \textit{users who are more pessimistic are more likely to establish negative relations than those with low level of pessimism}. Table~\ref{Tab:PersonalitytTest} summarizes the p-values for the above $t$-tests at significance level $\alpha =0.01$.

\begin{table}[!htbp]
	\footnotesize
	\vspace{-5pt}
	\centering
	\caption{\textbf{P-values of $t$-test results corresponding to the individuals personality theory analysis with $K=20$ at significance level $\alpha=0.01$}. The null hypothesis for each test is refuted.}\label{Tab:PersonalitytTest}
	\begin{tabular}{|l|l|l|}
		\hline
		& $\{{\bf h}_o,{\bf l}_o\}$ & $\{{\bf h}_p,{\bf l}_p\}$ \\ \hline
		Epinions &  $3.16\textrm{e}{-19}$ &  $5.36\textrm{e}{-10}$\\ \hline
		Slashdot &  $1.60\textrm{e}{-23}$ &  $7.18\textrm{e}{-12}$\\ \hline
	\end{tabular}
	\vspace{-10pt}
\end{table}

\subsubsection{Summary}
These results verify that established social psychological theories are applicable to social media data and in that they are correlated to the problem of signed link formation. Our findings could be summarized as follows:

\begin{itemize}[leftmargin=*]
	\item Users with higher positive (negative) emotion strengths are more likely to establish positive (negative) links than those with lower positive (negative) emotion strengths
	\item Users are more likely to follow their friends' behaviors in terms of positive and negative links.
	\item Users with high optimistic (pessimistic) behavior are more likely to establish positive (negative) links than those with low optimism (pessimism).
\end{itemize}

Findings from signed link analysis could benefit a variety of tasks of signed network mining as it introduces new sources of information which could alleviate the signed link sparsity problem. We next investigate how to incorporate these findings from social psychological theories as a guidance to feature engineering in addressing the data sparsity problem in signed networks.

\section{Social-Theory Guided Feature Engineering for Signed Link Prediction}
The previous section suggests that social psychological theories are applicable to the user-generated data and are also correlated with signed link formation. These theories could be then helpful in addressing the problem of sparse negative and positive links. We incorporate our findings into the task of link prediction in signed networks. Signed link prediction is an important problem in social computing as it can help to infer and understand attitude of a specific user toward other users using the extracted information from positive and negative links in the vicinity~\cite{leskovec2010predicting}. Signed link prediction also has many applications in network analysis tasks such as community detection and recommendation systems require information about positive and negative linkages between entities. For example, recommendation systems' outputs (i.e., product recommendation) are derived from other users' choices. Given negative linkage between users, the recommender can improve the quality of recommendation by avoiding to recommend items from one user's preferences to another.  Users with positive links are also more likely to share similar preferences~\cite{tang2016recommendations}. Another example of signed link prediction application is in political analysis that one needs to complete a picture of online political landscape. Inferring the signed links in political networks can further help description of rivalries and coalitions between groups~\cite{ozer2017negative}.

\begin{figure*}[t]\vspace{-15pt}
	\centering
	\subfloat[Postive link prediction]{\includegraphics[scale=0.37]{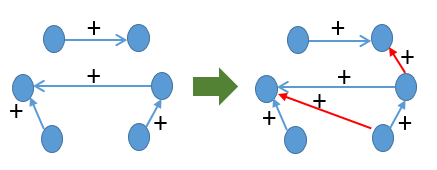}}\quad
	\subfloat[Sign prediction]{\includegraphics[scale=0.37]{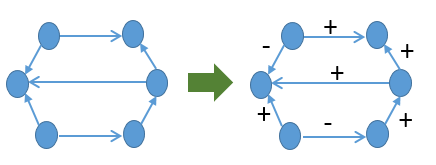}}\quad
	\subfloat[Signed link prediction]{\includegraphics[scale=0.37]{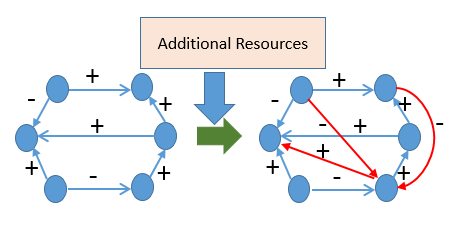}}
	\caption{\textbf{ Differences of positive link prediction, sign prediction and signed link prediction}}\label{diff}
	\vspace{-10pt}
\end{figure*}

The problem of signed link prediction is different and more challenging than its existing variants, positive link prediction~\cite{tang2013exploiting} and sign prediction~\cite{yang2012friend}. We illustrate the unique characteristics of the singed link prediction along with those of the existing variations, in Fig.~\ref{diff}. We list the key nuances of the signed link prediction problem as follows.
\begin{itemize}[leftmargin=*]
	\item In positive link prediction, we seek to predict only positive links from the existing ones. In contrast, as illustrated in Fig.~\ref{diff}(c), in signed link prediction, we aim to predict both positive and negative links simultaneously.
	\item Sign prediction problem infers the signs of the existing links (Fig.~\ref{diff}(b)). In the signed link prediction, we predict both link and its sign between users.
	\item A few factors can influence people in their establishment of positive links: since positive relations require time to nurture, a user usually does not have a large number of positive relations, or available explicit positive links are often sparse~\cite{tang2013exploiting}. Online users are often more willing to show positivity than negativity, therefore, negative links are much sparser than positive links in a signed network which results in the \emph{signed link sparsity} problem. Consequently, the problem of signed link prediction is more challenging than positive link prediction.	 
	\item The vast majority of existing works for the problem of signed link prediction~\cite{leskovec2010predicting}, aims at predicting positive and negative links by leveraging only the existing links between users. This could result in an inaccurate link formation due to the data sparsity problem. On the other hand, we seek to leverage additional resources such as user's emotional information and personality to alleviate the signed link sparsity problem.
\end{itemize}

\begin{wrapfigure}{R}{0.5\textwidth}
	\begin{center}
		\includegraphics[width=0.4\textwidth]{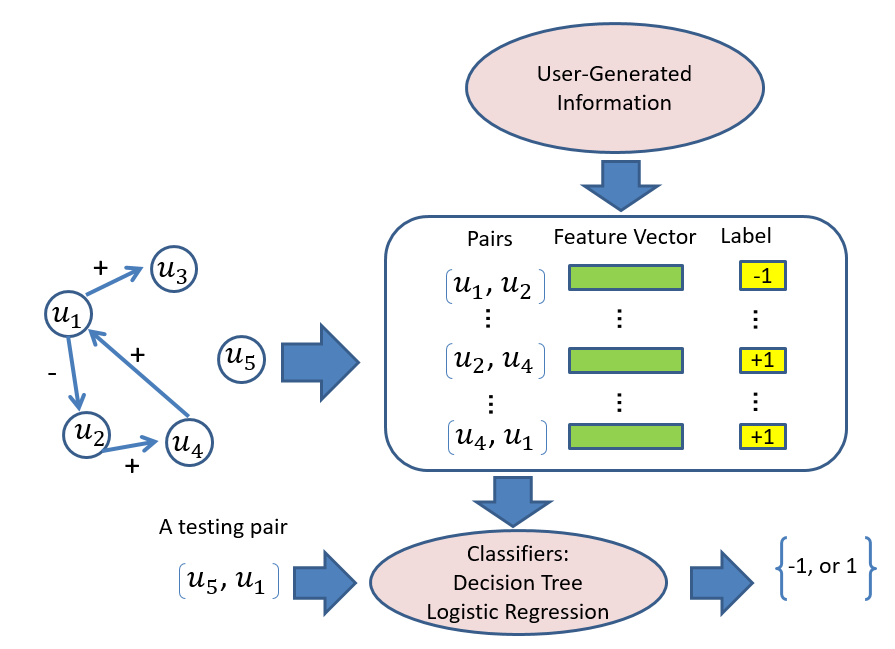}
	\end{center}
		\caption{\textbf{Positive and negative link predictor model based on network structure and other sources of user-generated information, i.e., personal and emotional information.}}\label{fig:LinkPrediction}
\end{wrapfigure}

Signed link prediction in signed social networks can be considered as a classification problem as shown in Fig.~\ref{fig:LinkPrediction} where (1) features are extracted from available sources to represent each pair of users; (2) existing positive links are considered as positive class with label $+1$, and existing negative links are considered as class with label $-1$
 ; and (3) a supervised classifier is trained by using these extracted features and labels. Given a pair of users $(u,v)$, we predict the signed link between them using the classifier probability estimation $p_c \in [0,1]$. Probability estimation represents the strength of the link between two users. If the probability estimation $p_c$ has a higher value than a given positive link threshold $\epsilon_p$, i.e., $p_c \geq \epsilon_p$, there is a positive link between $(u,v)$. If $p_c$ is less than a given negative link threshold $p_c < \epsilon_n$, there is a negative link between $(u,v)$. Note that $\epsilon_p \geq \epsilon_n$. Based on the classification framework for the link prediction problem, one effective way to incorporate findings from social psychological theories and link analysis is through feature engineering~\cite{leskovec2010predicting,dong2018feature}. This means, instead of brute force search for all possible features from available sources, we can extract features derived from our findings guided by social psychological theories. In particular, we extract three different categories of features which we discuss in details next.


The first category of social-theory-guided features is constructed with regard to the correlation between emotional information and signed link formation suggesting that (1) {\it with high probability, users with positive (negative) emotions are more likely to establish positive (negative) links than those without} and (2) \textit{users with higher positive (negative) emotion strengths are more likely to establish positive (negative) links than those with lower positive emotion strengths}. According to these findings, we introduce 6 features describing emotional information and their strengths, including the proportion of positive/negative emotions between a pair of users $i$ and $j$ and the proportion of positive/negative emotions between users $i$ and $j$ and the whole network.

To account for the findings of diffusion of innovation theory, we develop a group of 2 features which is in accordance with the theory that {\it users are likely to follow their friends' behaviors in terms of positive and negative link creation.} These features are based on the behaviors of user $u_i$'s friends (e.g. $u_k$) toward user $u_j$, including the proportion of friends $u_k$ who have established positive and negative links with user $u_j$.

Last but not least, we employ individual's personality theory which indicates the correlation between user's personality and positive and negative link formation. This leads to the introduction of the third group of 4 features from personality information for each pair of users $u_i$ and $u_j$, including optimism and pessimism scores for each user. Again, the introduced features align well with the theory, i.e., \textit{more optimistic (pessimistic) users are more likely to establish positive (negative) relations than those with low level of optimism (pessimism)}.

We have detailed how we capture our findings via social-theory-guided feature engineering for the problem of positive and negative link prediction in social networks. A summary of all introduced features is demonstrated in table~\ref{Tab:Features}. 
\begin{table*}[!htbp]
	\centering
	\footnotesize
	\caption{\textbf{Summary of group of features for a pair of users $(u_i$,$u_j)$}}
	\begin{tabular}{|p{2.5cm}|p{11cm}|}
		
		\hline
		Feature Group & Features\\\hline
		\multirow{6}{100pt}{EI: Emotional\\ information} & (1) Proportion of negative interactions between $(u_i,u_j)$ ($\frac{{\bf N}_{ij}}{{\bf N}_{ij}+{\bf P}_{ij}}$)\\ & (2) Proportion of positive interactions between $(u_i,u_j)$ ($\frac{{\bf P}_{ij}}{{\bf N}_{ij}+{\bf P}_{ij}}$)\\ & (3) Proportion of negative interactions of $u_i$ ($\frac{{\bf N}_{i}}{{\bf N}_{i}+{\bf P}_{i}}$)\\ & (4) Proportion of positive interactions of $u_i$ ($\frac{{\bf P}_{i}}{{\bf N}_{i}+{\bf P}_{i}}$)\\ & (5) Proportion of negative interactions of $u_j$ ($\frac{{\bf N}_{j}}{{\bf {\bf N}}_{j}+{\bf P}_{j}}$)\\ & (6) Proportion of positive interactions of $u_j$ ($\frac{{\bf N}_{j}}{{\bf N}_{j}+{\bf P}_{j}}$)\\
		\hline
		\multirow{2}{100pt}{DI: Diffusion of \\ innovation} & (1) Proportion of followees of $u_i$ whom established negative link to $u_j$ ($\frac{|\{u_k| {\bf F}(u_i, u_k)=1 \wedge {\bf F}(u_k,u_j)=-1\}|}{|\{u_k|{\bf F}(u_i,u_k)=1\}|}$) \\ & (2) Proportion of followees of $u_i$ whom established positive link to $u_j$ ($\frac{|\{u_k| {\bf F}(u_i, u_k)=1 \wedge {\bf F}(u_k,u_j)=1\}|}{|\{u_k|{\bf F}(u_i,u_k)=1\}|}$)\\
		\hline
		\multirow{4}{100pt}{IP: Individual's \\ personality} & (1) Optimism of $u_i$ \\ & (2) Pessimism of $u_i$ \\ & (3) Optimism of $u_j$ \\ & (4) Pessimism of $u_j$ \\
		\hline
	\end{tabular}
	
	\label{Tab:Features}\vspace{-10pt}
\end{table*}

\section{Experiments}

We first study how link analysis affects signed link prediction in signed social networks. Next, we examine the robustness of the signed link predictor against the data sparsity challenge. Then, we investigate the connection of the features to the theories. Finally, we perform feature importance analysis to examine how much each category of the features and their combinations contribute to the signed link prediction performance and how robust they are against the sparsity problem.

\subsection {Quality of Predicted Signed Links}
We train two classifiers, i.e., Decision Tree and $\it{l}_2$-Regularized Logistic Regression, on the set of features, to predict positive and negative links in the network (Fig.~\ref{diff}.c) . In all experiments, we use 10-fold cross validation, by treating one fold as the test set $\mathcal{S}$, and setting $\mathbf{F}_{ij}=0, \forall (u_i, u_j) \in \mathcal{S}$. Note that ${\bf F}_{ij} = 0$ indicates the case where the information regarding the link between users $u_i$ and $u_j$ is missing. We then feed new representation of ${\bf F}$ as the input to the predictor. We set the probability estimation threshold for positive and negative links as $\epsilon_p = \epsilon_n = 0.5$. This means that for a given pair of users, if the classifier probability estimation $p_c \geq 0.5$, there is a positive link between $(u,v)$ and the link is negative otherwise. We recall that $p_c$ demonstrates the link strength between two users. Therefore, we can consider a tighter bound for inferring signed links by increasing the difference between signed link thresholds $\epsilon_p$ and $\epsilon_c$. Different values of thresholds $\epsilon_p$ and $\epsilon_c$ can be examined in future.

Note that in the signed social networks, positive links are much denser than negative ones resulting in an imbalanced positive and negative links in both training and test sets. We employ different metrics to assess the performance of positive and negative link predictor on predicted values of links between pairs of users in the test set, $\{\widetilde{\mathbf{F}}_{ij} , \forall\langle u_i,u_j\rangle\in\mathcal{S}\}$. These metrics include,  accuracy (ACC), area under the curve (AUC)~\cite{fawcett2006introduction}, and individual precision (Precision+, Precision-), recall (Recall+, Recall-) and F1-value (F1+, F1-) for positive and negative links. We compare effectiveness of our approach against following representative positive and negative links predictors:
\begin{itemize}[leftmargin=*]
	\item {\bf All23}: This method~\cite{leskovec2010predicting} uses 23 different topological structure features of the network for each pair of link between users based on the local relations of a node and balance theory. The first group of features includes in-degree and out-degree numbers from positive and negative links and the second group exploit balance theory to extract triangle based features. We train the logistic regression and decision tree classifiers over these features.
    \item \textbf{VOTE}: This method~\cite{naaman2018edge} assigns a set of topological properties to each edge such as centrality measures (4 features), degrees (4 features), community measures (4 features) and scale motif frequencies (13 features). After normalizing each attribute, AdaBoost classifier is applied on features to predict positive/negative links for a given pair.
    \item \textbf{CTMS}: This method~\cite{khodadadi2017sign} uses tendency rate of triple-micro structures in signed networks. It first extracts closed triple micro structures around a pair of users assuming the edge between them is positive. It then repeats the same procedure assuming the edge between two users is negative. The edge sign is then determined by the scenario which has greatest tendency rate.
    \item \textbf{SC}: This method is a signed version of Laplacian matrix for signed network embedding. We choose the top-$d$ eigen vectors ($d$=20) corresponding to the smallest eigenvalues of the signed Laplacian matrix as the low dimensional node representations. Then, a logistic regression classifier is trained using nodes' embeddings to predict positive/negative links for a given pair of users.
    \item \textbf{DB/OP/RP}: This method~\cite{shahriari2016sign} first extracts two sets of topological-based features for each pair of users: the first set consists of seven (7) degree-based (DB) features, and the second set contains twelve (12) features describing user's optimism/reputation (OP/RP), which are derived from the links between users, a total of 19 features. Then, it trains a logistic regression classifier using these features to predict positive/negative links for a given pair of users.
	\item {\bf TDP}: This predictor treats positive and negative links prediction problem as the propagation of sequence of atomic operations~\cite{guha2004propagation}. In this method, positive link propagates multiple steps while negative relation propagates only a single step.
	\item {\bf Random}: This baseline randomly selects pairs of users with positive and negative links.
    \item \textbf{Majority}: This baseline assigns the sign with the majority number of edges to the given edge. All edges will be considered as positive as positive is the sign with the majority label in our datasets.
\end{itemize}

The purpose of this study is to investigate whether our signed link analysis can improve the link prediction performance. In this experiment, we focus on feature engineering to capture our observations. Note that there are other signed link predictor frameworks such as low rank matrix factorization, which could be deployed on the proposed features. In future, we plan to extend matrix factorization by generating constraints and defining objective functions.

{\it VOTE}, {\it CTMS}, {\it All23}, {\it SC} and {\it TDP} use topological network structure. Moreover, {\it DB/OP/RP} extracts users personality and reputation from network structure. We train both Decision Tree and $\it{l}_2$-Regularized Logistic Regression classifiers on the combination of our proposed features with those of {\it All23}--hereafter referred to as {\it All23+EI+DI+IP} method. 
Note that we do not consider the combination of our proposed features with {\it TDP} method, as it was already defeated by {\it All23}. The comparison results are summarized in Table~\ref{Tab:Res1} and we observe the followings:

\begin{table*} [t]  \centering \footnotesize
\caption{\textbf{Link prediction performance for Epinions and Slashdot. Metrics are Area Under the Curve (AUC), Accuracy (ACC), individual precision, recall and F1-value for positive/negative links.}}\label{Tab:Res1}
		\subfloat[\bf{Epinions}]{
				\begin{tabular}{@{} cl|cccccccc @{}}
					
					& & {\bf{{\footnotesize AUC}}} & {\bf{{\footnotesize ACC}}}& {\bf{{\footnotesize Precision+}}} & {\bf{{\footnotesize Precision-}}} & {\bf{{\footnotesize Recall+}}} & {\bf{{\footnotesize Recall-}}} & {\bf{{\footnotesize F1+}}} & {\bf{{\footnotesize F1-}}} \\
					\cmidrule{2-10}
                    & All23+EI+DI+IP (LR) & \textbf{0.8034} & \textbf{0.9451} & 0.9492 & 0.8195 & \textbf{0.9946} & 0.3377 &  \textbf{0.9713} & 0.4759 \\
					& All23+EI+DI+IP (DT)  & 0.7542 & 0.8897 & 0.9597 &  0.3637&  0.9195 & 0.5192 & 0.9387 & 0.4215\\
					& EI+DI+IP (LR)  & 0.7830 & 0.9422 & 0.9461 & 0.8023 & 0.9941 & 0.2944 & 0.9695 & 0.4307 \\
					& EI+DI+IP (DT) & 0.7417 & 0.9318 & 0.9648 & 0.5392 & 0.9614 & 0.5629 & 0.9631 & \textbf{0.5508}\\
					& All23 (LR) & 0.7830 & 0.9282 & 0.9331 & 0.5865 & 0.9936 & 0.1127 & 0.9624 & 0.1890\\
					& All23 (DT)  & 0.6554 & 0.7294 & \textbf{0.9682}  &  0.1886 & 0.7624  & \textbf{0.6885} & 0.8531 & 0.2961 \\
					& SC & 0.7682 & 0.8036 & 0.9012 & 0.7436 & 0.8741 & 0.2589 & 0.8874& 0.3841\\
                    
                    & VOTE & 0.7774 & 0.8873 & 0.8911 & \textbf{0.8677} & 0.8934 & 0.3118 & 0.8922&0.4587\\
                    
                    & CTMS & 0.7458 & 0.8612 & 0.8802 & 0.8444& 0.9271 & 0.2878 & 0.9030 & 0.4292\\
                    & DB/OP/RP & 0.5981 & 0.8979 & 0.9011 & 0.8218 & 0.9929 & 0.2034 & 0.9447& 0.3190\\

                    & TDP & 0.5682 & 0.9254 & 0.9289 & 0.4815& 0.955& 0.0510 & 0.9611 & 0.0922\\
                    & Random & 0.4998 &0.4999 & 0.9278 & 0.0722 & 0.5009 & 0.4952 & 0.6505 & 0.1260\\
                    
                    & Majority & 0.5 & 0.9273 & 0.9273 & 0& 1 & 0 & 0.9622 & 0 \\
					\cmidrule[1pt]{2-10}
				\end{tabular}}
			\quad
			\subfloat[\bf{Slashdot}]{	\begin{tabular}{@{} cl|cccccccc @{}}
					
					& & {\bf{{\footnotesize AUC}}} & {\bf{{\footnotesize ACC}}}& {\bf{{\footnotesize Precision+}}} & {\bf{{\footnotesize Precision-}}} & {\bf{{\footnotesize Recall+}}} & {\bf{{\footnotesize Recall-}}} & {\bf{{\footnotesize F1+}}} & {\bf{{\footnotesize F1-}}} \\
					\cmidrule{2-10}
                    & All23+EI+DI+IP (LR) & \textbf{0.9431} & \textbf{0.9412} & 0.9454& \textbf{0.9252} & 0.9796 & 0.8162& \textbf{0.9622} & \textbf{0.8673}\\
					& All23+EI+DI+IP (DT)  & 0.9029 & 0.9266 & 0.9550 & 0.8375& 0.9487& 0.8548& 0.9518& 0.8459\\
					& EI+DI+IP (LR)  & 0.9343 & 0.9175& 0.9248& 0.8880& 0.9711& 0.7436 & 0.9474 & 0.8094\\
					& EI+DI+IP (DT) & 0.8973 & 0.9343& \textbf{0.9597} & 0.8539& 0.9541& \textbf{0.8701} & 0.9569 & 0.8619\\
					& All23 (LR) & 0.8777 & 0.8370& 0.8677 & 0.6992& 0.9284& 0.5402 & 0.8970& 0.6094\\
					& All23 (DT)  & 0.8172 & 0.7994& 0.9399& 0.5487& 0.7881& 0.8364& 0.8573 & 0.6627\\
					& SC & 0.8219 & 0.8583 & 0.9236 & 0.7598 & 0.8925 & 0.3739 & 0.9077& 0.5011\\
					
                     & VOTE & 0.9106 & 0.9196 & 0.9370 & 0.8859& 0.9321& 0.6052 & 0.9345& 0.7191\\
                    
                     & CTMS & 0.8798 & 0.8925 & 0.9078 & 0.8614& 0.9286 & 0.5618 & 0.9180& 0.6800\\
                     & DB/OP/RP & 0.6347 & 0.9173 & 0.9246& 0.7506 & \textbf{0.9898} & 0.2527 & 0.9560 & 0.3781\\
                    
                    & TDP & 0.6826 & 0.9254 & 0.9289 & 0.4815& 0.9550 & 0.0510 & 0.9611 & 0.0922\\
                    & Random & 0.5024 & 0.4987 & 0.7231 & 0.2769 & 0.4980 & 0.5011 & 0.5898 & 0.3566\\
                    
                    & Majority & 0.5 & 0.7255 & 0.7255 & 0& 1 & 0 & 0.8409 & 0 \\
					\cmidrule[1pt]{2-10}
				\end{tabular}}\vspace{-25pt}

\end{table*}
\begin{itemize}[leftmargin=*]
	\item Decision tree and logistic regression classifiers have different learning biases which result in different performances as expected, while the logistic regression always achieve better performance. Results also show that high accuracy does not imply better performance in predicting both positive and negative links. In other words, some methods (e.g. All23, and TDP) have high accuracy but low recall and F1 for negative class. This confirms the effectiveness of AUC metric when we deal with highly imbalanced dataset.
	\item {\it All23} outperforms {\it Majority} and {\it Random} as well as {\it TDP}. The reason for improvement of {\it All23} over {\it TDP} is that edge signs could be leveraged in positive and negative link formation rather than requiring a notion of propagation from farther-off parts of the network as~\cite{guha2004propagation} did. {\it TDP} cannot also handle the problem of imbalance distribution of positive and negative links.
    \item {\it All23} outperforms {\it DB/OP/RP} but is inferior to {\it VOTE} and {\it CTMS} have better results than {\it All23}. The reason is that {\it VOTE} and {\it CTMS} consider the global trend and implicit forces that direct the sign of each relation.
    \item Performance results of {\it SC} is comparable to {\it All23} but cannot outperform {\it VOTE} and {\it CTMS} methods. The reason is that {\it VOTE} and {\it CTMS} explicitly incorporate the global network information while {\it SC} does not. In particular, {\it SC} seeks to capture network's properties which include the network sparsity and imbalance relationship between signed links. These properties are thus reflected in the learned nodes' representations and final signed predictions.
	\item We train both classifiers on the combination of our features {\it EI+DI+IP} and those of {\it All23}. For both datasets, {\it All23+EI+DI+IP} method outperforms other baselines, {\it All23}, {\it SC}, {\it VOTE}, {\it CTMS} and {\it DB/OP/RP}, using both logistic regression and decision tree classifiers. The reasons are twofold. First, the task of feature extraction based on the solely topological structure of signed networks may not be robust due to the sparsity of signed links, specifically negative links. Thus, there might be even many pairs of users without features based on balance theory~\cite{chiang2011exploiting}. Moreover, node embedding based approach {\it SC}, seeks to capture network's properties which include the network sparsity and imbalance relationship between signed links. These properties are thus reflected in the learned nodes' representations and lead to poor signed predictions. In contrast, {\it All23+EI+DI+IP} method considers auxiliary user information related features other than topological structure. Second, the imbalance problem of positive and negative links distribution cannot be handled by other approaches as well; while, exploiting additional resources in {\it All23+EI+DI+IP} mitigate the imbalance problem in link distribution.
    \item For both datasets and classifiers, {\it All23+EI+DI+IP} achieves better performance over {\it DB/OP/RP}, despite that both approaches leverage optimism/reputation-based features. The reason is {\it DB/OP/RP} uses topological structures to extract these features and hence suffers from the sparsity problem, similar to {\it All23}. Simply put, there could be many pairs of users with zero optimism/reputation, which make the optimism/reputation-based features less useful in alleviating the imbalance problem of signed links distributions. In contrast, {\it All23+EI+DI+IP} infers users' personality information from their feedback on different issues other than merely using signed links. {\it All23+EI+DI+IP} also exploits other sources of information inferred from non-structural source, i.e., emotional information, which can also help in addressing sparsity of signed links.
\item {\it All23+EI+DI+IP} and {\it EI+DI+IP} can better handle the problem of imbalance distribution of signed links in comparison to other approaches. This is because of leveraging additional resources of information which are not heavily imbalanced such as personality and emotional information.
\item For both datasets, all methods perform well in predicting labels for positive edges. The results of recall for negative class (i.e., Recall-) show that {\it All23+EI+DI+IP (LR)} has the best results in predicting negative edges amongst all methods. These results indicate that high performance in terms of accuracy (AUC), and F1+, does not necessarily indicate that the method can perform well for negative class. {\it All23+EI+DI+IP (LR)} has the best performance amongst all methods in terms of accurately predicting both positive and negative edges. This is because other methods rely solely on structural data which is both imbalanced and sparse while {\it All23+EI+DI+IP} leverages other non-structural sources of information for predicting signed links between users, i.e., emotional and personality information. These sources provide more information about negative interactions and possible future negative links between users and thus can handle the sparsity and imbalance challenge of signed link prediction.
	
    \item There is not a significant difference between the performance of {\it EI+DI+IP+All23} and {\it EI+DI+IP}. The reason is that features corresponding to {\it All23} are topological based and thus are not robust to signed link sparsity problem. Therefore adding them as additional features to {\it EI+DI+IP} does not make any significant improvement.	
	
\end{itemize}
Both of the two different classifiers achieved the same improvement when they were deployed on the proposed features. This verifies the significance of social psychological theories for signed link prediction. To recap, positive and negative link prediction based on additional user information, performs better than the representative signed link prediction approaches. Thus our findings in link analysis can significantly improve the performance of link prediction in signed social networks.
\subsection{Robustness to Data Sparsity}
As discussed earlier, signed networks suffer from data sparsity issue. 
	 Our goal is to leverage additional resources such as user's emotional information and personality to alleviate the signed link sparsity problem. Here, we examine how social-theory-guided engineered features are robust against data sparsity problem. Similar to the previous subsection, we use 10-fold cross validation for evaluation. Each time, we hold one fold out and treat it as our test set. From the remaining 9 folds, we pick $x\%$ of positive and $x\%$ of negative links to construct the training set. We vary $x$ as $\{60, 70, 80, 90, 100\}$ to investigate how well our signed link predictors perform with different sizes of training set and different sparsity scenarios, i.e., less training set is a sparser scenario. We report the AUC results in Table.~\ref{Tab:Res_sparsity} to assess the performance of predictors. We use AUC since it is more effective than other metrics when dealing with highly imbalanced dataset.

\begin{itemize}[leftmargin=*]
	\item In general, with the decrease of the training data, the performance of all methods deteriorates.
	\item {\it EI+DI+IP+All23} and {\it EI+DI+IP} are more robust against different sparsity scenarios in comparison to other approaches. Their performance dropped less than $5\%$ when $x$ decreases from $100\%$ to $60\%$, while performance of other signed link predictors decreases more than $10\%$ in terms of AUC. This is because other methods rely on topological features which get sparser by reducing the size of training set. However, {\it EI+DI+IP+All23} and {\it EI+DI+IP} leverage additional sources of non-topological information including personality and emotional related features which are not heavily sparse in comparison to structural network data. These sources provide more information in lack of enough signed network information and helps handling sparsity problem.
\end{itemize}

\begin{table*} [t]  \centering \footnotesize
	\vspace{-10pt}
	\caption{\textbf{Robustness of positive and negative link predictors to data sparsity problem for Epinions and Slashdot datasets. Area Under the Curve (AUC) is used as an evaluation metric.}}\label{Tab:Res_sparsity}
		\begin{tabular}{@{} cl|ccccc |ccccc@{}}
			& &  & & \bf{Epinions} &   & & & & \bf{Slashdot} & &  \\
			
			& & {\bf{{\footnotesize 100}}} & {\bf{{\footnotesize 90}}}& {\bf{{\footnotesize 80}}} & {\bf{{\footnotesize 70}}} & {\bf{{\footnotesize 60}}} &{\bf{{\footnotesize 100}}} & {\bf{{\footnotesize 90}}}& {\bf{{\footnotesize 80}}} & {\bf{{\footnotesize 70}}} & {\bf{{\footnotesize 60}}} \\
			\cmidrule{2-12}
			& All23+EI+DI+IP (LR) & \textbf{0.8034} & \textbf{0.7953} & \textbf{0.7880} & \textbf{0.7799} & \textbf{0.7705} & \textbf{0.9431} & \textbf{0.9386} & \textbf{0.9291} & \textbf{0.9217} & \textbf{0.9099}\\
			& All23+EI+DI+IP (DT)  & 0.7542 & 0.7426 & 0.7373 &  0.7301 &  0.7256  & 0.9029 & 0.8939 & 0.8859 &  0.8792 &  0.8710 \\
			& EI+DI+IP (LR)  & 0.7830 & 0.7722 & 0.7694 & 0.7618 & 0.7521 & 0.9343 & 0.9173 & 0.9089 & 0.8928 & 0.8793 \\
			& EI+DI+IP (DT) & 0.7417 & 0.7308 & 0.7225 & 0.7092 & 0.7014 & 0.8973 & 0.8893 & 0.8816 & 0.8700 & 0.8624  \\
			& All23 (LR) & 0.7830 & 0.7627 & 0.7514 & 0.7287 & 0.7099  & 0.8777 & 0.8582 & 0.8413 & 0.8163 & 0.7911\\
			& All23 (DT)  & 0.6554 & 0.6284 & 0.6066  &  0.5785 & 0.5711 & 0.8172 & 0.7924 & 0.7680  &  0.7504 & 0.7314  \\
			& SC  & 0.7682 & 0.7443 & 0.7267  &  0.7011 & 0.6925 & 0.8219 & 0.8064 & 0.7756  &  0.7639 & 0.7471  \\
			& VOTE & 0.7774 & 0.7619 & 0.7489 & 0.7321 & 0.7156  & 0.9106 & 0.8863 & 0.8711 & 0.8598 & 0.8439\\
			& CTMS & 0.7458 & 0.7197 & 0.7088 & 0.6944 & 0.6719 & 0.8798 & 0.8615 & 0.8523 & 0.8271 & 0.8037\\
			& DB/OP/RP & 0.5981 & 0.5651 & 0.5387 & 0.5118 & 0.5029 & 0.6347 & 0.6199 & 0.5916 & 0.5638 & 0.5329\\
			& TDP & 0.5682 & 0.5429 & 0.5179 & 0.4835& 0.4862 & 0.6826 & 0.6549 & 0.6278 & 0.6015& 0.5927\\
			\cmidrule[1pt]{2-12}
				\end{tabular}
		\end{table*}

\subsection{Connections of Features to the Theories}

As discussed earlier, our goal is to explore how different theories can be deployed in positive and negative link prediction. Here, we investigate how each social-theory-guided feature suggests evidence for the formation of positive/negative links. We study each group of features independently and follow the same procedure in~\cite{leskovec2010predicting} for analyzing balance and status theories. Using logistic regression classifier, we compute each feature's corresponding coefficient, odds ratio (OR) and $95\%$ confidence interval (CI). Signs of these coefficients indicate how the corresponding feature is used by logistic regression classifier and suggest an evidence for establishing either positive or negative links. In addition, the odds ratio value indicates a possible statistical relationship between the feature and signed links-- odd ratio greater than 1 suggests higher odds of \textit{positive link} for classifier outcome, while odd ratio less than 1 indicates higher odds of \textit{negative link}. The $95\%$ confidence interval is often used as a proxy for the presence of statistical significant association between the feature and expected  outcome (either positive link or negative link) if it does not overlap the null value, i.e. odds ratio value of 1. A small interval indicates a higher precision of odds ratio whereas a large interval indicates a low level of precision. These properties of logistic regression provide a natural connection between each feature and its corresponding theory.
\subsubsection{Emotional Information}\label{section:Emotional}
We train logistic regression classifier on emotional features. Tables~\ref{Tab:Emotional-Epinions} and ~\ref{Tab:Emotional-Slashdot} depict the results for Epinions and Slashdot data, respectively. The first column shows each feature number and the second column indicates the expected link sign based on the emotional information theory. As we see, the signs of all learned coefficients are the same as the expected signs for both data, thereby confirming the alignment of emotional information theory with the learned model. Odd ratios for features which are positively associated with positive expected degree have values greater than 1 whereas the values of features associated with negative expected degree are smaller than 1. The confidence interval for each feature does not contain the null value (OR=1) confirming the significant association between features and expected outcome. Moreover, small confidence intervals indicate the higher precision of odds ratio for each corresponding feature. These results show that proportion of positive/negative interactions between the pair $(u_i,u_j)$ as well as the proportion of positive/negative interactions source/target nodes have with others are good predictors of the signs of potential links $u_i$ and $u_j$ are going to make with others in future.
\begin{table}[t]
	\centering
	\footnotesize
	\caption{\textbf{Logistic regression coefficients based on emotional information features on Epinions}}\label{Tab:Emotional-Epinions}
	\begin{tabular}{c|c|c|c|c}
		Feature & Emotional Information & Coefficient & Standard Error & OR ($95\%$ Confidence Interval)  \\ \hline \hline
		1 & -1 & -4.7308 & 0.0084 & 0.0088 $\pm$ 0.009 \\
		2 & 1 & 1.2127 & 0.0024 & 3.3621 $\pm$ 3.38 \\
		3 & -1 & -3.0833 & 0.0156 & 0.0458 $\pm$ 0.44 \\
		4 & 1 & 0.8192 & 0.0053 & 2.2684 $\pm$ 2.30 \\
		5 & -1 & -3.0234 & 0.0156 & 0.0486 $\pm$ 0.051 \\
		6 & 1 & 0.7818 & 0.0053 & 2.1852 $\pm$ 2.21 \\\hline
		
	\end{tabular}
\end{table}

\begin{table}[t]
	\centering
	\footnotesize
	\caption{\textbf{Logistic regression coefficients based on emotional information features on Slashdot}}\label{Tab:Emotional-Slashdot}
	\begin{tabular}{c|c|c|c|c}
		Feature & Emotional Information & Coefficient & Standard Error & OR ($95\%$ Confidence Interval)  \\ \hline \hline
		1 & -1 & -3.8398 & 0.0036 & 0.0215 $\pm$ 0.022 \\
		2 & 1 & 2.0061 & 0.0020 & 7.4333 $\pm$ 7.47 \\ 
		3 & -1 & -0.7053 & 0.0081 & 0.4939 $\pm$ 0.51 \\
		4 & 1 & 2.3801 & 0.0032 & 10.8033 $\pm$ 10.88 \\ 
		5 & -1 & -4.2080 & 0.0050 & 0.0148 $\pm$ 0.015 \\ 
		6 & 1 & 1.3159 & 0.0033 & 3.7277 $\pm$ 3.76 \\\hline 
		
	\end{tabular}
\end{table}
\subsubsection{Diffusion of Innovation}
\begin{table}[t]
	\centering
    \footnotesize
	\caption{\textbf{Logistic regression coefficients based on diffusion of innovation features on Epinions}}\label{Tab:Followee-Epinions}
	\begin{tabular}{c|c|c|c|c}
		Feature & Diffusion of Innovation & Coefficient & Standard Error & OR ($95\%$ Confidence Interval) \\ \hline \hline
		1 & -1 & -0.6373 & 3.8354$e-04$ & 0.52903 $\pm$ 0.53 \\
		2 & 1 & 0.2396 & 3.3364$e-05$ &  1.2685 $\pm$ 1.27 \\\hline 
	\end{tabular}
\end{table}
\begin{table}[t]
	\centering
    \footnotesize
	\caption{\textbf{Logistic regression coefficients based on diffusion of innovation features on Slashdot}}\label{Tab:Followee-Slashdot}
	\begin{tabular}{c|c|c|c|c}
		Feature & Diffusion of Innovation & Coefficient & Standard Error & OR ($95\%$ Confidence Interval) \\ \hline \hline
		1 & -1 & -0.1744 & 1.7605$e-04$ & 0.8396 $\pm$ 0.84 \\ 
		2 & 1 & 0.0212 & 1.3364$e-05$ & 1.02132 $\pm$ 1.022 \\\hline
	\end{tabular}
\end{table}
\begin{table}[t]
	\centering
    \footnotesize
	\caption{\textbf{Logistic regression coefficients for individual personality features on Epinions}}\label{Tab:IP-Epinions}
	\begin{tabular}{c|c|c|c|c}
		Feature & Individual Personality & Coefficient & Standard Error & OR ($95\%$ Confidence Interval)  \\ \hline \hline
		1 & 1 & 1.2567 & 0.0137 & 3.5133 $\pm$ 3.61 \\
		2 & -1 & -0.5146 & 0.0104 & 0.5977 $\pm$ 0.62 \\
		3 & 1 & 1.6042 & 0.0113 & 4.9732 $\pm$ 5.09 \\
		4 & -1 & -0.7118 & 0.0111 & 0.4907 $\pm$ 0.51 \\\hline 
	\end{tabular}
\end{table}
\begin{table}[t]
	\centering
    \footnotesize
	\caption{\textbf{Logistic regression coefficients for individual personality features on Slashdot}}\label{Tab:IP-Slashdot}
	\begin{tabular}{c|c|c|c|c}
		Feature & Individual Personality & Coefficient & Standard Error & OR ($95\%$ Confidence Interval)  \\ \hline \hline
		1 & 1 & 0.3397 & 0.0098 & 1.4045 $\pm$ 1.44 \\
		2 & -1 & -0.1170 & 0.0033 & 0.8895 $\pm$ 0.90 \\
		3 & 1 & 0.3538 & 0.0098 & 1.4245 $\pm$ 1.46 \\ 
		4 & -1 & -0.1262 & 0.0033 & 0.8814 $\pm$ 0.89 \\\hline
	\end{tabular}
\end{table}
Following the same procedure of what we have done for emotional information, we consider features of diffusion of innovation group. The learned coefficients for this problem are shown in Tables~\ref{Tab:Followee-Epinions} and ~\ref{Tab:Followee-Slashdot} for Epinions and Slashdot, respectively. The signs of coefficients are exactly the same as what we expected according to the theory of {\it diffusion of innovation}, discussed earlier. The odd ratio for feature number 1 is smaller than 1 which confirms the feature's association with expected negative sign. The odd ratio for second feature is also greater than 1 which indicates the positive association of corresponding feature with expected positive sign. The confidence interval for both feature does not contain the null value (OR=1) confirming the significant association between features and expected outcome. These results confirm that the behavior of $u_i$'s followees toward $u_j$ are good predictors of the signs of potential links between the pair $(u_i,u_j)$, and further demonstrate that these features align well with this theory.
\subsubsection{Individual's Personality}
Results of training logistic regression on four features of this group are depicted in Tables~\ref{Tab:IP-Epinions} and \ref{Tab:IP-Slashdot} which confirm the alignment of individual's personality theory with the learned model. Both first and third features have the positive coefficients with the odd ratio greater than 1. Also, the confidence interval does not include null value which further confirms the positive significant association of these features with potential positive expected link between the pair $(u_i,u_j)$. Coefficient of second and last features are negative. Moreover, their odd ratio values are smaller than 1 along with confidence interval less than 1. These results suggest the alignment of pessimism related features with the potential negative expected link between $(u_i,u_j)$.
\subsection{Feature Importance Analysis}
We discuss how important our different sets of the features are, in predicting positive and negative links in signed networks. We follow the same procedure in~\cite{leskovec2010predicting} and train two classifiers, random forest (with 200 trees) and logistic regression, over each category of features independently. We also train a decision tree classifier on the datasets and it achieves the worst performance. Therefore, for brevity we only report feature analysis and classification results when random forest and logistic regression are used. Results using 10-fold cross validation are shown in Tables \ref{Tab:Importance1} and \ref{Tab:Importance2}. 
\begin{table}[t]
	\centering
    \footnotesize
	\caption{\textbf{Performance for analyzing the importance of different combinations of features for Epinions. The metrics are AUC, ACC and F1-score for individual positive/negative classes.}}\label{Tab:Importance1}
    \begin{tabular}{@{} cl|cccc|cccc @{}}
    &  {\bf{{ Feature Group}}} &  \multicolumn{4}{c}{\textbf{Logistic Regression}}& \multicolumn{4}{c}{\textbf{Random Forest}} \\[1ex]
   &  & {\bf{{\footnotesize AUC}}} & {\bf{{\footnotesize ACC}}}& {\bf{{\footnotesize F1+}}} & {\bf{{\footnotesize F1-}}} & {\bf{{\footnotesize AUC}}} & {\bf{{\footnotesize ACC}}} & {\bf{{\footnotesize F1+}}} & {\bf{{\footnotesize F1-}}} \\
					\hline\hline
		& Emotional Information (EI) & \textbf{0.7478}  &0.9227 &0.9583 & \textbf{0.4048}& \textbf{0.7314} &\textbf{0.9611}& \textbf{0.9721} & \textbf{0.5622}\\
		& Diffusion of Innovation (DI) & 0.7390 & \textbf{0.9304} &\textbf{0.9636} & 0.3479& 0.7146 & 0.9314 &0.9688 &0.4952\\	
		& Individual Personality (IP) & 0.6702 & 0.9254& 0.9611& 0.3887 &0.6315 & 0.9411 & 0.9624 &0.4385\\
		& All23 & 0.7830 &0.9282 &0.9624 &0.1890 & 0.6777&0.7385 &0.8716 &0.3254 \\
		& EI+DI & 0.7842 & 0.9325 & 0.9635& 0.5538& 0.7586 &0.9316 &0.9714 &0.4398\\	
		& EI+IP & 0.7554 & 0.9376&0.9673 &0.5212 & 0.7327 &0.9318 & 0.9718 &0.4889\\
		& DI+IP & 0.7162 & 0.9303 & 0.9636&0.1966 & 0.7216&0.9326 &0.9684 &0.4516\\	
		& EI+DI+IP & 0.7830 & 0.9422& 0.9695& 0.4307&  0.7624 &0.9508 &0.9831 &0.5718\\	
		& All23+EI+DI+IP  & 0.8034 & 0.9451& 0.9713& 0.4759 & 0.7812 &0.9007 &0.9517 &0.5974 \\	\hline
	\end{tabular}
\end{table}

\begin{table}[t]
	\centering
    \footnotesize
	\caption{\textbf{Performance for analyzing the importance of different combinations of features for Slashdot. The metrics are AUC, ACC and F1-score for individual positive/negative classes.}}\label{Tab:Importance2}
    \begin{tabular}{@{} cl|cccc|cccc @{}}
    &  {\bf{{ Feature Group}}} &  \multicolumn{4}{c}{\textbf{Logistic Regression}}& \multicolumn{4}{c}{\textbf{Random Forest}} \\[1ex]
   &  & {\bf{{\footnotesize AUC}}} & {\bf{{\footnotesize ACC}}}& {\bf{{\footnotesize F1+}}} & {\bf{{\footnotesize F1-}}} & {\bf{{\footnotesize AUC}}} & {\bf{{\footnotesize ACC}}} & {\bf{{\footnotesize F1+}}} & {\bf{{\footnotesize F1-}}} \\
					\hline\hline
		& Emotional Information (EI) & \textbf{0.9085}  & 0.9186& \textbf{0.9337}& \textbf{0.7406} & \textbf{0.8694} & \textbf{0.9183} & \textbf{0.9473} & \textbf{0.8174}\\
		& Diffusion of Innovation (DI) & 0.8953 & \textbf{0.9628} & 0.9066 &0.5673 & 0.8321 &0.8752 &0.92014 & 0.7174 \\	
		& Individual Personality (IP) & 0.7248 & 0.8704& 0.9155 &0.7218 &0.8229 & 0.8889 & 0.9288 & 0.7475\\
		& All23 & 0.8777 &0.8370 &0.8970 &0.6094 & 0.8306 &0.8479 &0.8966&0.7119 \\
		& EI+DI & 0.9249 & 0.9271 & 0.9522& 0.8464& 0.9141 &0.9480  &0.9664 &0.8850 \\	
		& EI+IP & 0.9128 & 0.9414& 0.9474 & 0.8108& 0.8953 &0.9371 &0.9595 & 0.8594 \\
		& DI+IP & 0.8972 & 0.9635 & 0.9747 & 0.5650 & 0.8883 & 0.9315 & 0.9558 & 0.8472\\	
		& EI+DI+IP & 0.9343 &0.9175& 0.9474& 0.8094&  0.9151 & 0.9411 & 0.9671& 0.8873\\	
		& All23+EI+DI+IP  & 0.9431 & 0.9412 & 0.9622 & 0.8673 & 0.9277 & 0.9584 &0.9731 &0.9078\\	\hline
        
	\end{tabular}
\end{table}

Logistic regression for the datasets using the emotional information features achieves AUC scores of $0.7478$ and $0.9085$. This shows that these features are the most effective ones. They also achieve the highest F1-scores indicating the efficiency of emotional information in predicting negative links even when the dataset is highly imbalanced and sparse, specifically for negative links. The next important feature is the feature of user's friends' behavior, while the individual personality is the least effective one. Another interesting observation is that the combination of all three groups of features performs slightly better than the combination of each pair of features. This suggests that each feature though different from others, contribute almost equally to the signed link analysis problem. Finally, the combination of three features performs better than {\it All23}. This verifies that merely using topological information is not sufficient for signed link analysis due to the sparsity of links; while auxiliary information can mitigate the sparsity problem in signed networks.
\section{Conclusion and Future Work}
In this work, we study how findings from social psychological theories can help mitigate the problem of signed link analysis by using auxiliary user information. This is specifically important as signed link data is often sparse. Our unique contribution lies at the use of social psychological theories to exploit characteristics of the signed networks rather than topological features. These theories guide us to use auxiliary user information such as user's personality and emotional information which are available in the form of users' opinions, likes, and postings. In particular, we employ \textit{Emotional Information}, \textit{Diffusion of Innovations}, and \textit{Individual Personality}, for link analysis in signed networks by extracting three different categories of features. We further demonstrate the connection of these theories to the signed link prediction problem. Extensive experiments in signed link prediction show the significance of the social-theory-guided features for signed link analysis. Our results show the importance of each feature as well as its role in data sparsity problem in signed link analysis.

In future, we would like to study the evolution of users' emotions and personality traits over time and investigate if dynamic signed networks can benefit from deploying these theories. This could be done by exploiting the changes of the features over time in dynamic signed link prediction~\cite{chen2018exploiting}. Furthermore, insights from our work is directly useful in studying the growth of signed social networks and constructing synthetic signed networks that reflect our findings from social/psychological theories. We also plan to 
 extend this work to address the signed link sparsity problem in presence of partial and asymmetrical information. 
\begin{acks}
This material is based upon the work supported, in part, by NSF \#1614576, ARO W911NF-15-1-0328 and ONR N00014-17-1-2605.
\end{acks}
\bibliographystyle{ACM-Reference-Format}

\end{document}